\documentclass[twocolumn,amsmath,amssymb,superscriptaddress,nofootinbib,11pt,a4paper]{revtex4}

\pdfoutput=1
\usepackage[T1]{fontenc}
\usepackage{CJK}
\usepackage{xcolor}
\usepackage{amsfonts}
\usepackage{epstopdf}
\usepackage{wrapfig} 

\usepackage{graphicx}  
\usepackage{dcolumn}   
\usepackage{bm}
\usepackage{float}
\usepackage[T1]{fontenc}
\usepackage{CJK}
\usepackage{xcolor}
\usepackage{amsfonts}
\usepackage{epstopdf}
\usepackage{wrapfig} 

\usepackage{dcolumn}   
\usepackage{bm}
\usepackage{float}
\usepackage{braket}

\usepackage{subcaption}
\usepackage{mathrsfs}

\begin{document}

\title{Constraints on Covariant Horava-Lifshitz Gravity from precision measurement of planetary gravitomagnetic field}

\author{Li-dong Zhang}
\affiliation{\footnotesize\begin{tabular}[t]{@{}l@{}}School of Physical Science and Technology, Lanzhou University, Lanzhou 730000, China\end{tabular}}

\author{Li-Fang Li}\email{lilifang@imech.ac.cn}
\affiliation{\footnotesize\begin{tabular}[t]{@{}l@{}} Center for Gravitational Wave Experiment, National Microgravity Laboratory, Institute of \\
Mechanics, Chinese Academy of Sciences, Beijing 100190, China\end{tabular}}

\author{Peng Xu}
\affiliation{\footnotesize\begin{tabular}[t]{@{}l@{}} Center for Gravitational Wave Experiment, National Microgravity Laboratory, Institute of \\
Mechanics, Chinese Academy of Sciences, Beijing 100190, China\end{tabular}}
\affiliation{\footnotesize\begin{tabular}[t]{@{}l@{}}Lanzhou Center of Theoretical Physics, Lanzhou University, Lanzhou 730000, China\end{tabular}}
\affiliation{\footnotesize\begin{tabular}[t]{@{}l@{}} Taiji Laboratory for Gravitational Wave Universe (Beijing/Hanzhou), University of Chinese\\
Academy of Sciences, Beijing 100049, China\end{tabular}}
\affiliation{\footnotesize\begin{tabular}[t]{@{}l@{}}Hangzhou Institute for Advanced Study, University of Chinese Academy of Sciences,Hangzhou\\
310024, China\end{tabular}}

\author{Xing Bian}
\affiliation{\footnotesize\begin{tabular}[t]{@{}l@{}} Center for Gravitational Wave Experiment, National Microgravity Laboratory, Institute of \\
Mechanics, Chinese Academy of Sciences, Beijing 100190, China\end{tabular}}

\author{Ziren Luo}
\affiliation{\footnotesize\begin{tabular}[t]{@{}l@{}} Center for Gravitational Wave Experiment, National Microgravity Laboratory, Institute of \\
Mechanics, Chinese Academy of Sciences, Beijing 100190, China\end{tabular}}
\affiliation{\footnotesize\begin{tabular}[t]{@{}l@{}} Laboratory of Gravitational Wave Precision Measurement of Zhejiang Province, Hangzhou\\
Institute for Advanced Study, UCAS, Hangzhou 310024, China.\end{tabular}}

\begin{abstract}
{As a generalization of Einstein's theory, Horava-Lifshitz has attracted significant interests due to its healthy ultraviolet behavior. In this paper, we analyze the impact of the Horava-Lifshitz corrections on the gravitomagnetic field. We propose a new planetary gravitomagnetic field measurement method with the help of the space-based laser interferometry, which is further used to constrain the Horava-Lifshitz parameters. Our analysis shows that the high-precision laser gradiometers can indeed limit the parameters in Horava-Lifshitz gravity and improve the results by one or two orders when compared with the existing theories. Our novel method provides insights into constraining the parameters in the modified gravitational theory, which facilitates a deeper understanding of this complex framework and paving the way for potential technological advancements in the field.}
\end{abstract}


\maketitle

\onecolumngrid
\noindent \textbf{Keywords:}Horava-Lifshitz Gravity ; Gravitomagnetism; gravitational constant; Gravitational gradient measurement; Space-based Experiments;  Parametrized Post-Newtonian

\newpage

\twocolumngrid
 
\section{Introduction}

Quantum mechanics(QM) and Einstein’s general relativity (GR) are two major revolutionary developments in physics during the early part of the 20th century. However, the incompatibility of these two theories has posed significant challenges for modern physics. Combining them to formulate a theory of quantum gravity remains elusive due to numerous difficulties such as non-renormalizability and the nature of spacetime at small scales. Among a few candidate theories existing, Horava-Lifshitz gravity has attracted significant attentions~\cite{Horava2009,Blas2009}.

The Horava-Lifshitz gravity proposed in 2009~\cite{Horava2009} modifies the Einstein-Hilbert action by introducing higher-order spatial derivative terms. The primary motivation behind the Horava-Lifshitz gravity is to construct a theory of quantum gravity that is renormalizable in the ultraviolet (UV) regime and 
 resolves the limitations of standard GR. Meanwhile, we also expect that this theory is not renormalizable and breaks down at very high energies where quantum effects become significant. Based on the above considerations, Horava and Melby-Thompson proposed a covariant version of Horava-Lifshitz gravity by introducing an extra local U(1) symmetry in 2010, which resolved the previous undesirable properties, including infrared instability~\cite{Horava2009,Wang2010} and strong coupling~\cite{I.Kimpton2010,Charmousis2009,K.Koyama2010,Wang2011}. This covariant Horava-Lifshitz gravity is consistency with cosmology~\cite{Huang2012,Huang2012non}. Additionally, the tests in the solar system have shown consistency with observations when the gauge field and the Newtonian prepotential are included in the metric~\cite{Lin2012}.

On the other hand, there are many parameters in the Horava-Lifshitz gravity. But when we consider the solar system, the terms from the cosmological constant and the space curvature are negligible. And the solution depends only on a few parameters. Explicitly when we expand this covariant Horava-Lifshitz gravity to the second post-Newtonian order, the potentials depend on the gravitational constant \(G\), the coefficient of the extrinsic curvature term \(\lambda\) characterizing deviations of the kinetic part of the action from GR and the two arbitrary coupling constants \(a_1\) and \(a_2\) in the matter action. To constrain these parameters, we use the solar system experiment tests within the Parametrized Post-Newtonian (PPN) framework. The PPN is a theoretical framework for a parameterized approach that approximates the effects of gravitational fields in various theories of gravity by introducing parameterized post-Newtonian corrections~\cite{Poisson2014}. With the help of this PPN formalism, all the parameters present in Horava-Lifshitz gravity are expressed in terms of PPN parameters~\cite{Will2018}. And we can constrain them with such solar system experiment.

As we know, a recent research focus is to use the gravitomagnetic effects to limit the parameters in  different gravity models. Within the weak field approximation and the slow motion limit, the dynamics of the linearized Einstein field equations may be formally identified with that of Maxwell's equations~\cite{Forward1961,Thorne1988,Maartens1998,Mashhoon2003}. The certain off-diagonal elements of the spacetime metric can be identified as the gauge potential in Maxwell's theory, thereby defining the gravitational analog of the magnetic field. Gravitomagnetism(GM) is an entirely relativistic effect, and its measurement provides an experimental test of the gravity theory on a planetary scale. In this paper we focus on the Lense-Thirring (Frame-dragging) effects and aim to constrain the above mentioned Horava-Lifshitz parameters~\cite{Lense1918}.

So far, the experimental measurements of such physical effects have directly constrained the parameters of the Horava-Lifshitz theory~\cite{Ciufolini1996}. A well-known  Gravity Probe B (GP-B) experiment~\cite{GP-B} measured the frame-dragging effect, constraining the difference between the gravitational constant \(G\) in Horava-Lifshitz theory and the Newtonian gravitational constant \(G_N\), yielding \(\left|\frac{G}{G_N}-1\right| \leq 0.05\) and \(\left| \frac{2}{3}\left(\frac{G}{G_N}a_1-\frac{a_2}{a_1}-1\right) \right|  \leq 0.0006\). Additionally, the LAGEOS satellites (Laser Geodynamics Satellites), launched by NASA (LAGEOS) and NASA-ASI (LAGEOS-2)~\cite{Lageos}, have tested the Lense-Thirring effect. With the laser-ranging techniques to measure distances and the combination of the two LAGEOS nodal longitudes that eliminate the uncertainty in the value of the Earth's quadrupole moment, this mission improved sensitivity and constrained the parameter to \(\left|\frac{G}{G_N}-1\right|  \textcolor{red}{\leq} 0.006\). Furthermore, the Monte Carlo simulations~\cite{Ciufolini2013} constrained the combination of the Horava-Lifshitz gravitational constant \(G\) and the Newtonian gravitational constant \(G_N\) to differ from unity by \(2 \times 10^{-3}\). Unfortunately, for the generic planetary sources, the gravitomagnetic effects are many orders lower than the Newtonian ones, which presents a challenge to the high precision measurement in the experiment.

\begin{figure}[!htbp]
    \centering    
    \includegraphics[height=2.2cm]{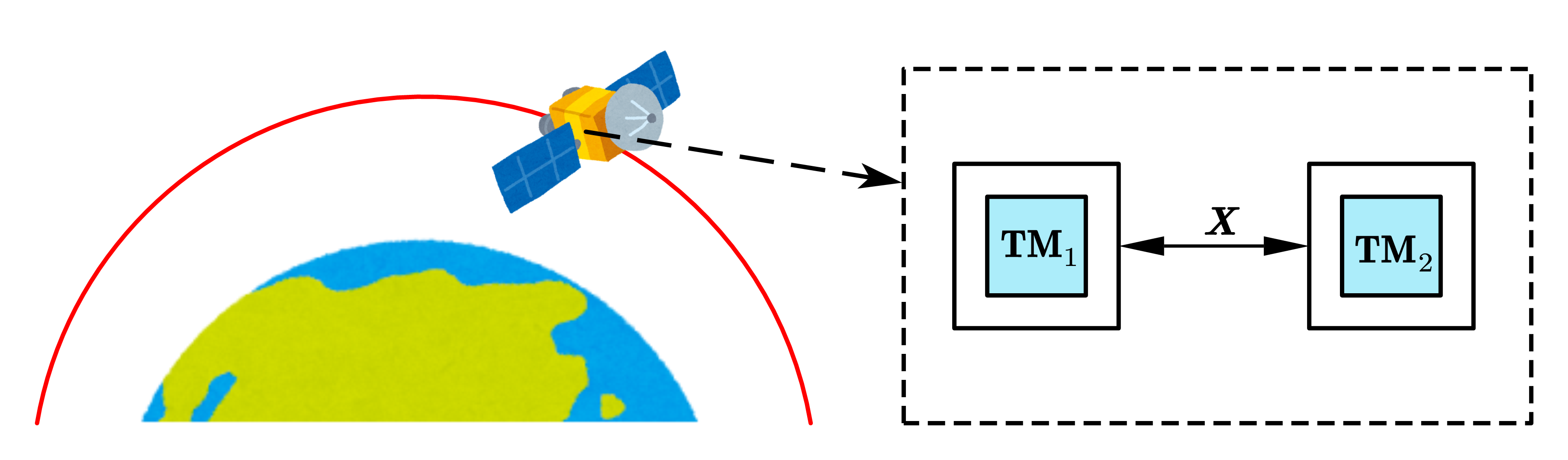}
    \caption{A scheme to measure gravitational gradients using two test masses on satellites in orbit}
    \label{S/C and TMs}
\end{figure}

Considering the above difficulties and challenges, our present work aims to establish a new theoretical framework for differential measurement of the planetary GM field with a space-based gradiometer. Unlike the standard readout of a gradiometer 
in terms of differential acceleration between the two test masses (TMs)~\cite{Moody2002,Rummel2011}, this new design helps us improve the detection precision. In theory, the GM field is a part of the spacetime curvature and it will produce the tidal forces acting on freely-falling test masses(TMs). Therefore the tidal force generated by the GM field can be measured directly in terms of satellite gradiometry. As we know, the two free-falling TMs in the LISA PathFinder mission~\cite{lisamcnamara2006,lisaarmano2015} naturally constitute a 1-dimensional gravity gradiometer. It is natural to wonder whether such kind of mission is also capable of measuring the GM field around a planet. Our studies show that by tracking the relative displacement of the two TMs in the direction transverse to the orbital plane(see Fig.~\ref{S/C and TMs}), we are able to read out the GM field of the Earth by differential measurement and test the classical solar system experiments. The basic schematic diagram is shown in~Fig.~\ref{S/C and TMs}. During the on-orbit science phase, we will track their relative motions generated by the GM tidal force with on-board laser interferometry. Explicitly, at the second post-Newtonian level, the two test masses located 50 cm apart and orbited the Earth governed by the Clohessy-Wiltshire equations derived from the geodesic deviation. Over a shorter time than the Lense-Thirring precession period (\(\sim 10^7\) years) of the orbital plane, the tidal force will generate a forced oscillation between the two test masses in the direction transverse to the orbital plane to realize the differential measurement and measure the Earth's GM field. 
The paper is organized as follows. We will briefly review the covariant Horava-Lifshitz gravity in Sec.\ref{sec2}. In Sec.\ref{sec3}, we solve the geodesic deviation equation up to the second post-Newtonian level for the two TMs located in the along-track direction of an almost circular orbit. We identify the measurable signals generated by the GM field in Sec.\ref{sec3}. In Sec.\ref{sec4}, we discuss the method of this space-based gradient measurement and constrain the corresponding parameters of such covariant Horava-Lifshitz gravity through this satellite experiment. Finally, we provide conclusions and discussions in the last Sec.\ref{sec5}.

\section{The gravitomagnetism of the covariant Horava-lifshitz gravity and its measurement schemes}
\label{sec2}
In this paper, we study the Lense-Thirring effect in the covariant Horava-Lifshitz gravity. It extends the symmetry of the Horava-Lifshitz gravity by introducing an additional U(1) gauge field and scalar field. Then a generalized covariant theory of gravity is obtained which eliminates the extra degrees of freedom ``spin-0 gravitons". And most importantly, it provides a general method for coupling gravity to other fields such as the matter field, which may cause Horava-Lifshitz gravity to differ from general relativity in the IR. Here we focus on this new version which includes the coupling \(\lambda\) in the extrinsic curvature term of the action~\cite{Silva2011,Horava2010}. Explicitly, the gauge field \(A(t,x)\) and the Newtonian pre-potential \(\phi(t,x)\) have been introduced to solve the scalar graviton problem. And more, this theory satisfies the projectability condition, meaning the lapse function only depends on time, \(N = N(t)\). Taking all these into consideration, the total gravitational action is given by~\cite{Horava2010,Silva2011}

\begin{widetext}
\begin{equation}\label{HLaction}
S_g=\zeta^2\int dt\  d^3x\  N\sqrt{g}\left(\mathcal{L}_K-\mathcal{L}_V+\mathcal{L}_\phi+\mathcal{L}_A\right),
\end{equation}
\end{widetext}
where $g=\text{det}(g_{ij})$ and
\begin{eqnarray*}
\mathcal{L}_K&=&K_{ij} K^{ij} -\lambda K^2,\\
\mathcal{L}_\phi&=&\phi\ \mathcal{G}^{ij}\left(2K_{ij}+\nabla_i\nabla_j \phi\right),\\
\mathcal{L}_A&=&\frac{A}{N}\left(2 \Lambda_g-R\right),
\end{eqnarray*}
where \(\mathcal{L}_K\) is the kinetic energy term, \(\mathcal{L}_\phi\) and \(\mathcal{L}_A\) are extra field terms. Here the covariant derivatives as well as the Ricci terms are all refer to the 3-metric \( g_{ij} \). The extrinsic curvature is represented by \( K_{ij} = g_i^k \nabla_k n_j \), where \( n_j \) is a unit normal vector to the spatial hypersurface. \(K\) is the trace of \(K_{ij}\). The 3-dimensional generalized Einstein tensor is denoted by \( \mathcal{G}_{ij} = R_{ij} - \frac{1}{2} g_{ij} R + \Lambda_g g_{ij} \). We recognize \(R\) as the linearized Ricci scalar of \(g_{ij}\).$\Lambda_g$ is the space curvature. The potential term \( \mathcal{L}_V \) refers to the potential part of the Lagrangian density~\cite{Horava2010,Silva2011}.

Matter coupling generalizes a scalar-tensor extension of the theory~\cite{Lin2012,Lin2014}, allowing the necessary coupling to emerge in the IR without compromising the power-counting renormalizability of the theory in the ultraviolet (UV). Specifically, the matter action term is given by~\cite{Horava2010,Silva2011}
\begin{equation}
S_M=\int dt d^3 x\tilde{N}\sqrt{\tilde{g}}\  \mathcal{L}_M\  (\tilde{N}, \tilde{N}_i,\tilde{g}_{ij}; \psi_n),
\end{equation}
where \(\mathcal{L}_M\) is the matter Lagrangian, and \(\psi_n\) represents the matter fields. The metric and the matter fields couple to the Arnowitt-Deser-Misner (ADM) components \(\tilde{N},\tilde{N}_i,\tilde{g}_{ij}\), as defined in~\cite{Horava2010,Silva2011,Lin2014}.

In the post-Newtonian approximations, we assume that the metric can be written in the form~\cite{Will2018}
\begin{eqnarray}
\gamma_{\mu\nu}=\eta_{\mu\nu}+h_{\mu\nu},
\end{eqnarray}
where $\eta_{\mu\nu}=diag(-1,1,1,1)$, and
\begin{eqnarray}
\label{perturbation_h}
h_{00}&\sim& \mathcal{O}(2)+\mathcal{O}(4)\nonumber,\\
h_{0i}&\sim&\mathcal{O}(3)\nonumber,\\
h_{ij}&\sim&\mathcal{O}(2)+\mathcal{O}(4),
\end{eqnarray}
where \(\mathcal{O}(n) \equiv \mathcal{O}(v^n)\). It should be noted that, in contrast to GR, \(h_{ij}\) needs to be expanded to the fourth order in \(v\) to obtain consistent field equations for the Hamiltonian constraint,  the momentum constraint and the trace part of the dynamical equations.

Based on the perturbation discussed in Eq.(\ref{perturbation_h}), given in~\cite{Will2018}

\begin{eqnarray}
\label{PPN}
h_{00} &\sim& 2U + \mathcal{O}(4) \nonumber,\\
h_{0i} &\sim& k V_i + f \chi_{,0i} + \mathcal{O}(5) \nonumber,\\
h_{ij} &\sim& 2 \gamma U \delta_{ij} + \mathcal{O}(4),
\end{eqnarray}
where the gauge freedom has been used to eliminate anisotropic terms in the space-space contribution of the perturbation. The coefficients are solved at the appropriate order which are as follows:

\begin{eqnarray}
\label{coeff}
k&=&-4\frac{G}{G_N}\nonumber,\\
f&=&\frac{G}{G_N}\frac{2-a_1-\lambda(4-3a_1)}{2(1-\lambda)}\nonumber,\\
\gamma&=&\frac{G}{G_N} a_1-\frac{a_2}{a_1},
\end{eqnarray}
where \(G_N\) is the Newton constant. \(G_N\) is in principle different from \(G\). \(a_1\) and \(a_2\) are two arbitrary coupling constants in the matter action.

\begin{figure}[!htbp]
    \centering    
    \includegraphics[height=6.5cm]{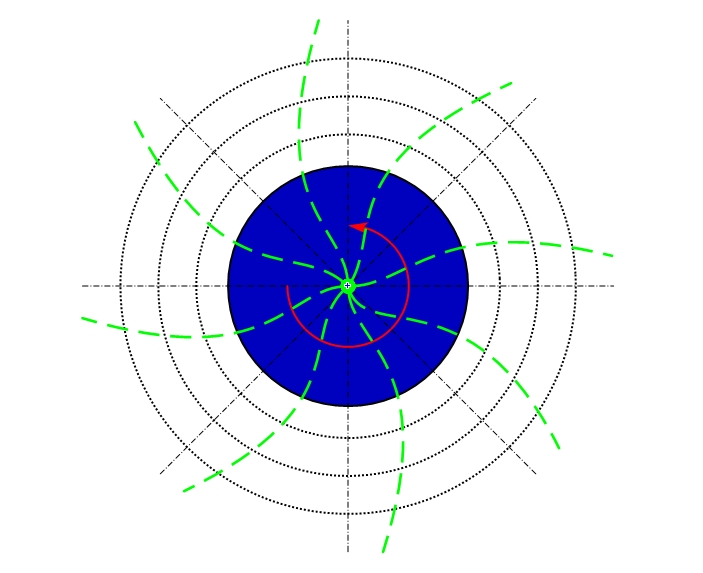}
    \caption{Diagram of Frame-dragging effect.}
    \label{Drageffect1}
\end{figure}

\begin{figure}[!htbp]
    \centering    
    \includegraphics[height=6.2cm]{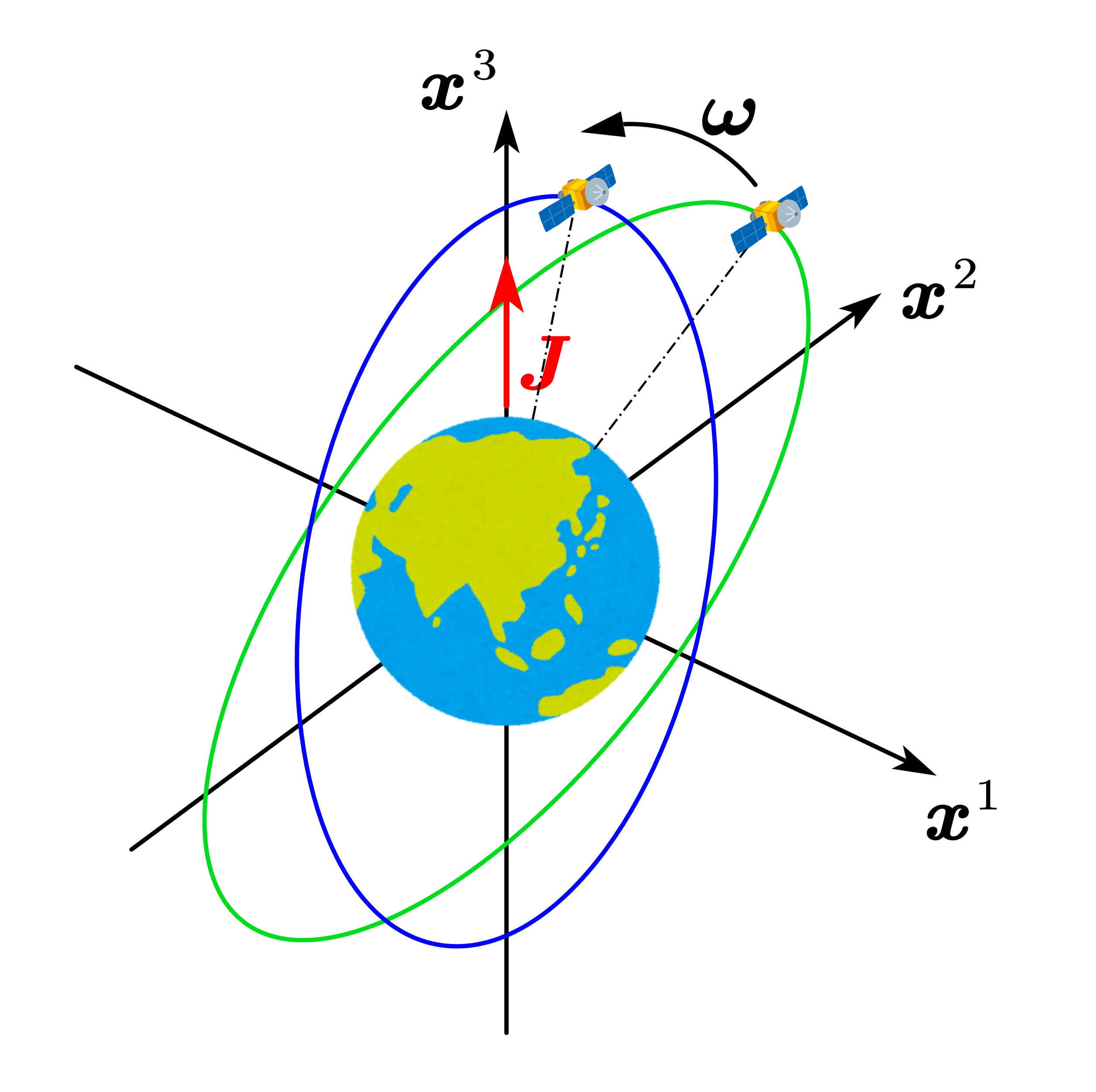}
    \caption{The impact on the general satellite orbit by the Frame-dragging effect.}
    \label{Drageffect2}
\end{figure}

Before we delve into the tedious post-Newtonian calculations, let us first describe the physical picture underlying the proposed measurement scheme. We consider the simple case of a nearly circular orbit for the spacecraft (S/C) and choose the Earth-centered inertial frame as the reference frame for simplicity. The presence of the GM field will generate the Lense-Thirring precession of the orbital plane about the Earth's rotation axis (see Fig.~\ref{Drageffect1} and Fig.\ref{Drageffect2}). The precession rate of the orbital normal \(N\) relative to the Earth-centered inertial frame is given by~\cite{Lense1918}

\begin{equation}
\Omega^{N} = \frac{2 G J \sin i}{G_N a^3},
\end{equation}
where \(i\) is the orbital inclination. Simultaneously, the GM field will also generate a precession of the orthonormal frame \(E_{(a)}^{i}\) attached to the S/C~\cite{Schiff1960}, which will precess about the Earth's rotation axis with a different angular rate with respect to the Earth-centered inertial frame, given by
\begin{equation}
\Omega^{S/C} = \frac{G J \sin i}{2 G_N a^3}.
\end{equation}

Therefore, by regarding the two TMs as markers of the orbit, the projection of the position difference \(Z^i\) in the transverse direction will measure the difference in precession. This constant offset between these two precessing rates will give rise to a relative oscillation of the two TMs along the transverse direction of the S/C (see Fig.\ref{frame1} and Fig.\ref{frame2}). Thus, the dominant GM signal in the S/C transverse direction \(E_{(3)}^{i}\) grows as 
\begin{align}
s_{GM} &\sim& d \sin(\Omega^N t - \Omega^{S/C} t) \sin(\omega t) \nonumber \\
&\sim& \frac{3 d G J t \sin i \sin(\omega t)}{2 G_N a^3},
\end{align}
where \(\omega\) is the orbital frequency. Due to this differential GM signal, the uncertainty (measurement error) in determining the globally fixed reference system will not be relevant to the proposed experiment.

By measuring the relative displacement of the TMs in the transverse direction using the on-board laser interferometry, the frame-dragging precession can be tracked very precisely. This approach is distinct from the LAGEOS or LARES missions, which attempt to track the orbital plane precession of a satellite through variations in the Keplerian elements with respect to the geocentric frame. Furthermore, as the GM precession is with respect to a globally defined inertial reference, it measures the differential precession of two gyroscopes: the rolling S/C along the orbit and the orbit of the S/C itself. This differential measurement will avoid many technical challenges in determining a global reference frame, such as those encountered in the GP-B experiment. Next, we will provide the detailed derivations of the second order post-Newtonian geodesic equations for the relative motion between the free-falling TMs in the S/C local frame.
\section{Derivation of the gravitomagnetic signal}
\label{sec3}

Based on the above discussions, the two freely-falling TMs in the along-track direction naturally form a one-dimensional gravity gradiometer. This setup allows us to measure the relative acceleration of the two TMs induced by the tidal gravitational force acting between them. To proceed with our calculations, we need to specify the orbit and the local tetrad shown in Fig.\ref{frame1} and Fig.\ref{frame2}. Here, we will consider a polar orbit as an example. Specifically, the 1PN (first order post-Newtonian) approximation for a nearly circular (spherical) orbit can be derived as follows
\begin{align}
&x=a \cos \omega\tau \cos \frac{2 G J}{G_N a^3}\tau-a \cos i \sin \omega\tau \sin\frac{2 G J}{G_N a^3}\tau,\nonumber \\
&y=a \cos i\sin \omega\tau \cos \frac{2 G J}{G_N a^3}\tau+a \cos \omega\tau \sin \frac{2 G J}{G_N a^3}\tau,\nonumber \\
&z=a\sin i\sin \omega\tau.
\end{align}

The initial longitude of the ascending node is zero, and the true anomaly is given by \(\Psi = \omega \tau\), where \(a\) denotes the orbit radius and \(\omega\) represents the mean angular frequency relative to the proper time along the orbit. And the 1PN local tetrad \(E_{(a)}\ ^{i}\) up to the first order post-Newtonian shown in Fig.\ref{frame1} and Fig.\ref{frame2} can be derived as follows~\cite{Xu2019}

\begin{widetext}
\begin{align}
&E^\mu_0=\begin{pmatrix}1 + a^2 \omega^2/2 + \gamma M/a\\
-a \omega \sin\omega\tau-\frac{2 \eta J \cos i (\omega\tau\cos\omega\tau+\sin\omega\tau)}{a^2}\\a\omega\cos i \cos \omega\tau+\frac{2 \eta J (\cos \omega\tau-\omega\tau\sin \omega\tau)}{a^2}\\
a \omega\sin i \cos\omega\tau
\end{pmatrix},\nonumber\\
&E^\mu_1=\begin{pmatrix}(a+2M)\omega\\
-(1+\frac{a^2\omega^2}{2}-\gamma\frac{M}{a})\sin\omega \tau+\frac{\eta J \omega\tau \cos i\cos \omega\tau}{a^3\omega}\\
(1+\frac{a^2 \omega^2}{2}-\gamma\frac{M}{a})\cos i \cos \omega\tau+\frac{\eta J \omega \tau \sin \omega\tau (1+3\cos 2i)}{4 a^3 \omega}\\
(1+\frac{a^2\omega^2}{2}-\gamma\frac{M}{a})\sin i\cos\omega\tau+\frac{3 \eta J \omega \tau \sin 2i \sin \omega \tau}{4 a^3 \omega}
\end{pmatrix},\nonumber\\
&E^\mu_2=\begin{pmatrix}-\frac{3 \eta J \omega \tau \cos i}{a^2}\\
(1-\gamma\frac{M}{a})\cos\omega \tau+\frac{\eta J \omega\tau \cos i\sin \omega\tau}{a^3\omega}\\
(1-\gamma\frac{M}{a})\cos i \sin \omega\tau-\frac{\eta J(\omega\tau\cos\omega \tau(1+3\cos 2i)+6 sin^2 i\sin\omega \tau)}{4 a^3 \omega}\\
(1-\gamma\frac{M}{a})\sin i\sin\omega\tau-\frac{3 \eta J \sin 2i (\omega\tau\cos \omega\tau-\sin \omega\tau)}{4 a^3 \omega}
\end{pmatrix},\nonumber\\
&E^\mu_3=\begin{pmatrix}-\frac{3 \eta J \omega\tau \sin i \sin\omega\tau}{2 a^2}\\
\frac{\eta J\sin i (3\sin(2\omega\tau)-2\omega\tau)}{4 a^3 \omega}\\
(1-\gamma\frac{M}{a})\sin i+\frac{3 \eta J \sin i \cos i \sin^2 \omega\tau}{2 a^3\omega}\\
-(1-\gamma\frac{M}{a})\cos i+\frac{3 \eta J\sin^2 i\sin^2\omega\tau}{2 a^3 \omega}
\end{pmatrix}
\end{align}
\end{widetext}
 where $\eta=\frac{G}{G_N}=-\frac{k}{4}$. With the local tetrad $E_{(a)}^{i}$ in hand, the geodesic deviation is explicitly expanded as

\begin{widetext}
    
\begin{figure}[!htbp]
    \centering    
    \includegraphics[height=7.5cm]{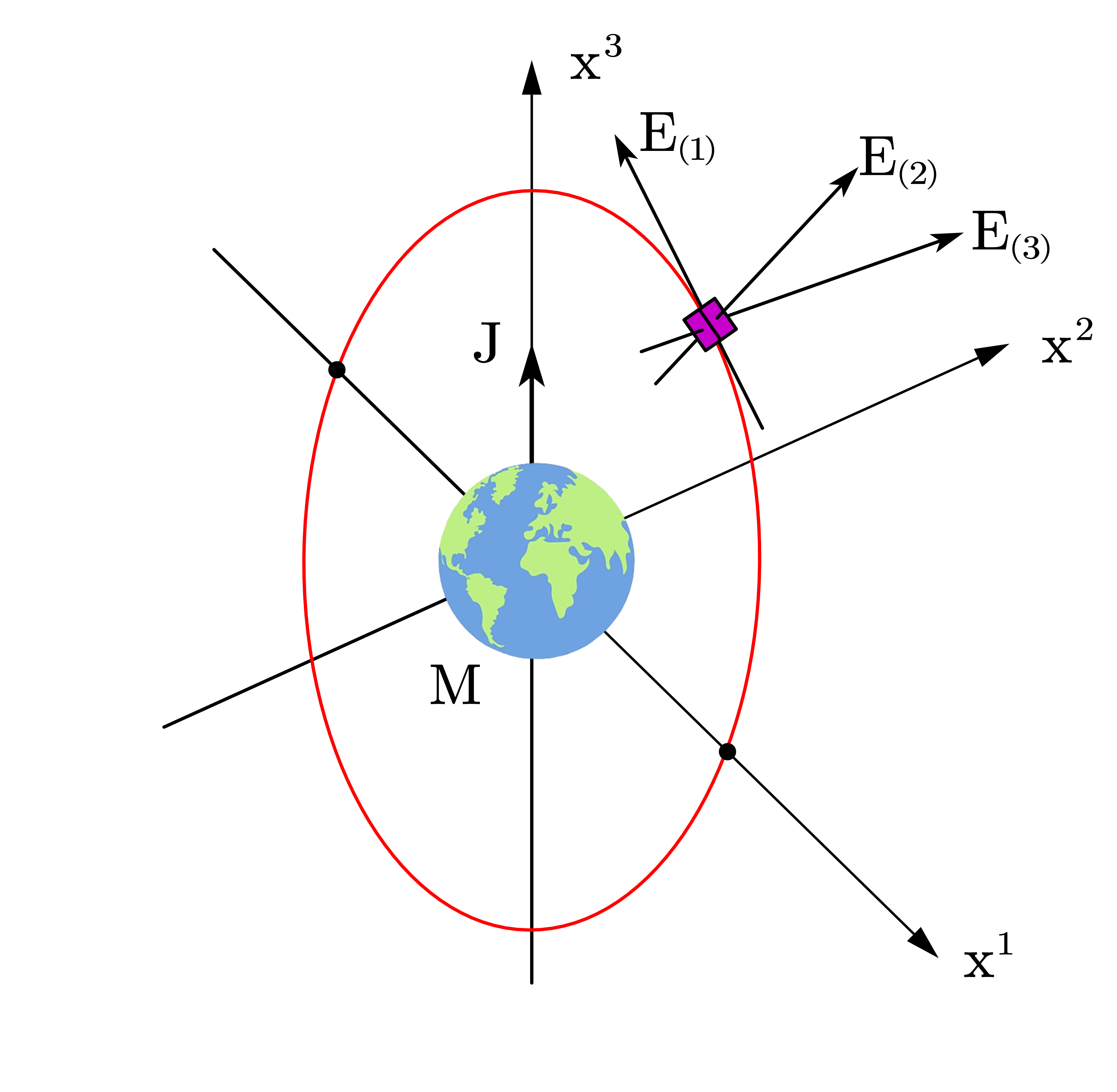}
    \caption{The LT precession of the polar nearly circular orbit.}
    \label{frame1}
\end{figure}

\begin{figure}[!htbp]
    \centering    
    \includegraphics[height=6cm]{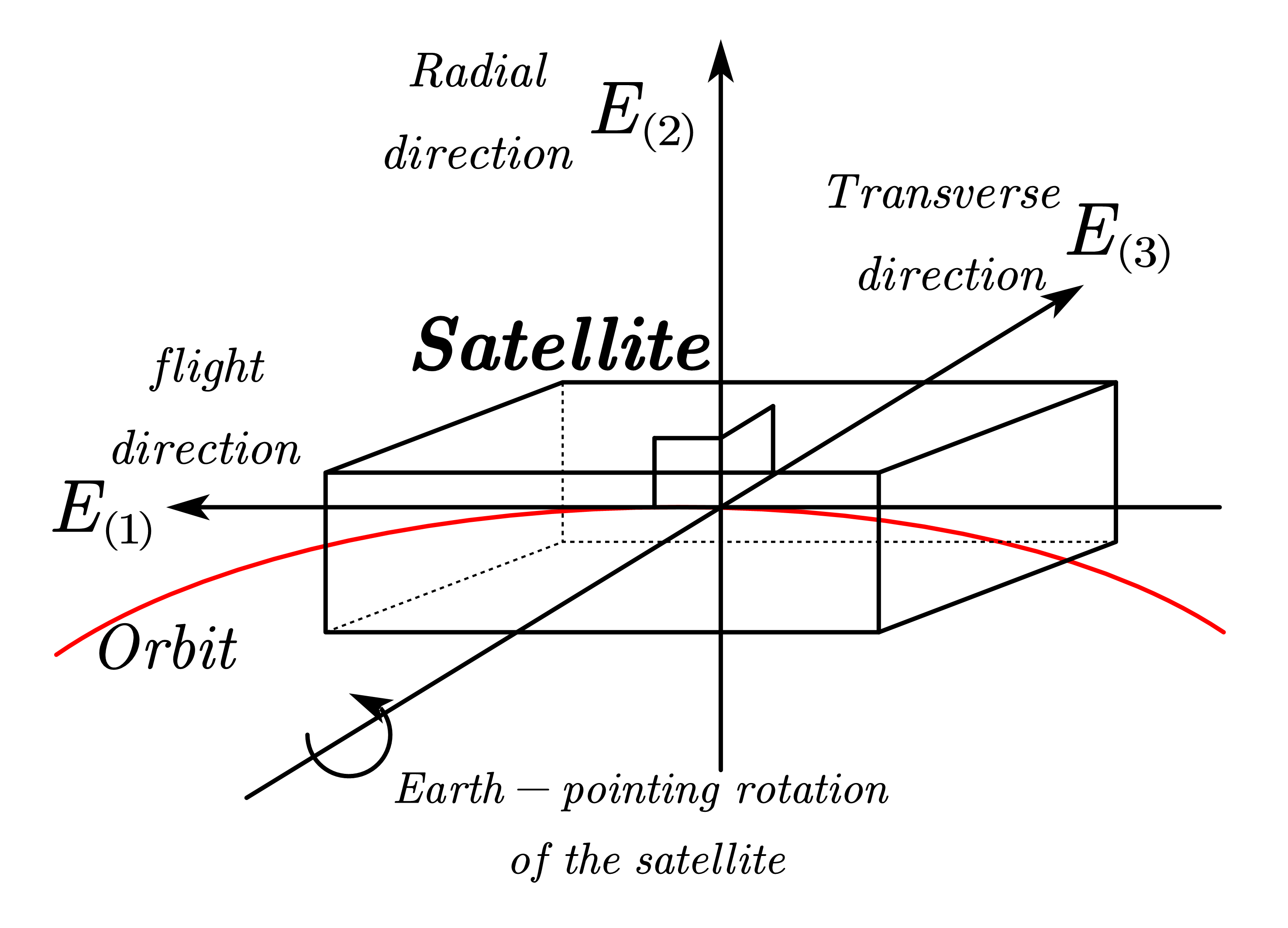}
    \caption{The local frame The Earth pointing orientation of the satellite~\cite{Xu2019}}
    \label{frame2}
\end{figure}

\begin{eqnarray}
\label{geodesic}
\frac{d^2 }{d \tau^2} X^{(a)}=-2\gamma^{(a)}_{ \ (b)(0)}\frac{d}{d\tau} X^{(b)}
-(\frac{d}{d\tau}\gamma^{(a)}_{ \ (b)(0)}+\gamma^{(c)}_{ \ (b)(0)}\gamma^{(a)}_{ \ (c)(0)})X^{(b)}
-K_{(b)}^{ \ (a)}X^{(b)},
\end{eqnarray}

\end{widetext}
where \( X^i=X^{(a)} E_{(a)}^i \) and \( \gamma^{(a)}_{\ (b)(c)}=E^{(a)}(\bigtriangledown_i E_{(b)j})E^i_{(c)} \) are the Ricci rotation coefficients. The first two terms in the above equation represent the Coriolis and inertial tidal forces, originating respectively from the relative rotation of the spacecraft's local frame with respect to the parallel propagated frames. The last term accounts for the tidal force generated by spacetime curvature.

With all the results in place, we substitute the relevant quantities into the geodesic deviation Eq.(\ref{geodesic}). By retaining the 1PN terms and neglecting contributions beyond \( \frac{X^{(a)}}{a^2} \mathscr{O}(\epsilon^4) \) and \( \frac{X^{(a)} \Psi}{a^2} \mathscr{O}(\epsilon^4) \) $(\epsilon=\frac{M}{r}$ is about $10^{-5}\sim 10^{-6}$), and after some algebraic manipulations, the time component of the geodesic deviation equation in the spacecraft's local frame turns out to be trivial, as expected.

\begin{widetext}
\begin{align}
\ddot{X}^{(0)}(\tau)=0,
\end{align}
\end{widetext}
and the spital parts may be expressed in a rather elegant form as:

\begin{widetext}
\begin{align}
\label{equation_1_1}
        \ddot{X}^{(1)}(\tau)+\omega(2-3a^2\omega^2)\dot{X}^{2}(\tau) 
        &+\frac{(\gamma-1)a^5\omega^4+6\eta J\omega \cos{i}}{a^3}X^{(1)}(\tau) 
        -\frac{9\eta J \tau \omega^2 \cos{i}}{a^3}X^{(2)}(\tau) \nonumber \\
    &+\frac{3\eta J\omega\sin{i} \cos{\omega \tau}}{a^3}X^{(3)}(\tau)=0 
\end{align}

\begin{align}
\label{equation_1_2}
    \ddot{X}^{(2)}(\tau)-(2\omega-3a^2\omega^3)\dot{X}^{(1)}(\tau) 
    &-\frac{9\eta J\tau\omega^2\cos{i}}{a^3}X^{(1)}(\tau) 
    +[2a^2(2+\gamma)\omega^4-3\omega^2 \nonumber \\ 
    &-\frac{6\eta J\omega\cos{i}}{a^3}]X^{(2)}(\tau) 
    -\frac{9\eta J \omega\sin{i}(\omega\tau\cos{\omega\tau}+3\sin{\omega\tau})}{2a^3}X^{(3)}(\tau)=0
\end{align}

\begin{align}
\label{equation_1_3}
    \ddot{X}^{(3)}(\tau)+\frac{3\eta J \omega\sin{i}\cos{\omega\tau}}{a^3}X^{(1)}(\tau) 
    &-\frac{9\eta J \omega\sin{i}(\omega\tau\cos{\omega\tau}+3\sin{\omega\tau})}{2a^3}X^{(2)}(\tau) \nonumber \\
    &+[\omega^2+a^2\omega^4(\gamma-1)]X^{(3)}(\tau) =0
\end{align}
\end{widetext}

The above equations are the basis for our next measurement. In the proposed measurement scheme, two TMs are positioned along the along-track direction with a separation distance \( d \sim 50 \) cm and follow an almost circular orbit of radius \( a \). Therefore, we can assume the initial conditions with the slight misalignments and deviations from the along-track orientation as

\begin{align}
& \frac{X_0^{(1)}}{d} =-1+\mathscr{O}(\lambda),\\
&\frac{X_{0}^{(2)}}{d } \sim\frac{X_{0}^{(3)}}{d }\sim\frac{\dot{X}_{0}^{(a)}}{d \omega}\sim\mathscr{O}(\lambda)<<1.
\end{align}

Under these conditions, the Eqs.(\ref{equation_1_1})-(\ref{equation_1_3}) are further simplified as
\begin{widetext}
\begin{eqnarray}
\ddot{X}^{(1)}_{PN}(\tau)+2\omega \dot{X}^{(2)}_{PN}(\tau)-d (\gamma-1)a^2\omega^4 
-\frac{6 \eta d J \omega\cos i}{a^3}=0\\
\ddot{X}^{(2)}_{PN}(\tau)-2\omega \dot{X}^{(1)}_{PN}(\tau)-3\omega^2{X}^{(2)}_{PN}(\tau) 
+\frac{9 \eta d J \tau \omega^2 \cos i}{a^3}=0\\
\ddot{X}^{(3)}_{PN}(\tau)+\omega^2{X}^{(3)}_{PN}(\tau) 
-\frac{3 \eta d  J \omega\sin i \cos(\omega\tau)}{a^3}=0.
\end{eqnarray}
\end{widetext}

These equations agree well with the physical picture of a transverse forced harmonic oscillator. Solving the above differential equations, we obtained the signals we needed as follows

\begin{widetext}
\begin{eqnarray}
&X^{(1)}_{PN}(\tau)=\frac{12 \eta d J \cos (i) \sin ^2\left(\frac{\omega\tau}{2}\right)}{a^3 \omega }+d \mathscr{O}(\epsilon^2\lambda),
\label{result-signal-1}
\\
&X^{(2)}_{PN}(\tau)=\frac{3 \eta d J \cos (i) (\tau  \omega -\sin (\omega\tau))}{a^3 \omega }+d \mathscr{O}(\epsilon^2\lambda),
\label{result-signal-2}
\\
&X^{(3)}_{PN}(\tau)=\frac{3 \eta d J \tau  \sin (i) \sin (\omega\tau)}{2 a^3}+d \mathscr{O}(\epsilon^2\lambda).
\label{result-signal-3}
\end{eqnarray}
\end{widetext}

The above equations (\ref{result-signal-1})-(\ref{result-signal-3}) are the signal that we're going to measure using a high-precision gravity gradient measurement mission. The amplitude of oscillation will grow linearly in time which is able to measure by the superconducting gradiometers. And from the perspective of frame-dragging precessions, such growing oscillation is generated by the differential precession of the spacecraft orientation and the orbital plane about the Earth's rotation axis. This is in contrast to the GP-B mission and the LAGEOS/LARES experiments which attempted to measure frame-dragging effects with respect to a certain globally defined reference frame. And more, this differential precession method improves the measurement accuracy which will be discussed in detail in the next section.

\section{Constrains from space experiments}
\label{sec4}
Currently, two highly promising schemes exist for enhancing the accuracy of gravity gradient measurements: the superconducting gravity gradiometer and the laser gravity gradiometer. Both the superconducting gravity gradiometer and the laser gravity gradiometer represent cutting-edge technologies with distinct advantages in enhancing the accuracy and precision of gravity gradient measurements. Each approach leverages advanced principles of physics and engineering to overcome traditional measurement limitations, making them invaluable tools in modern scientific research and applications.

Explicitly, the superconducting gravity gradiometer utilizes advanced superconducting technology to assess gravity gradients. At its core is the superconducting accelerometer, which operates based on the magnetic interaction between a current-carrying coil and a superconductor in a fully diamagnetic Meissner state. This setup forms a magnetic spring oscillator that leverages the Meissner effect, zero-resistance effect, and Josephson macroscopic quantum effect of superconductors to create a highly sensitive micro-displacement detection unit. One of the key advantages of the superconducting gravity gradiometer is its extremely low internal noise, which surpasses the measurement resolution limits of traditional room-temperature instruments.

On the other hand, the laser gravity gradiometer relies on the principle of dual-path interference. It determines the rate of change of gravity across three-dimensional space by employing two laser interferometric absolute gravimeters arranged differentially. This configuration allows the instrument to measure the gravitational acceleration of two falling bodies relative to their respective reference points using laser interferometry. By analyzing the positional shifts of these bodies, the instrument derives the gravity gradient at the measurement points through differential calculations. The laser gravity gradiometer's differential measurement approach helps to mitigate measurement errors caused by vibrations, thereby significantly enhancing measurement precision. Moreover, with the slight adjustments to the optical path and the position of the falling bodies, the system can be adapted to measure transverse gravity gradients.

\begin{figure}[!htbp]
    \centering    
    \includegraphics[height=3.5cm]{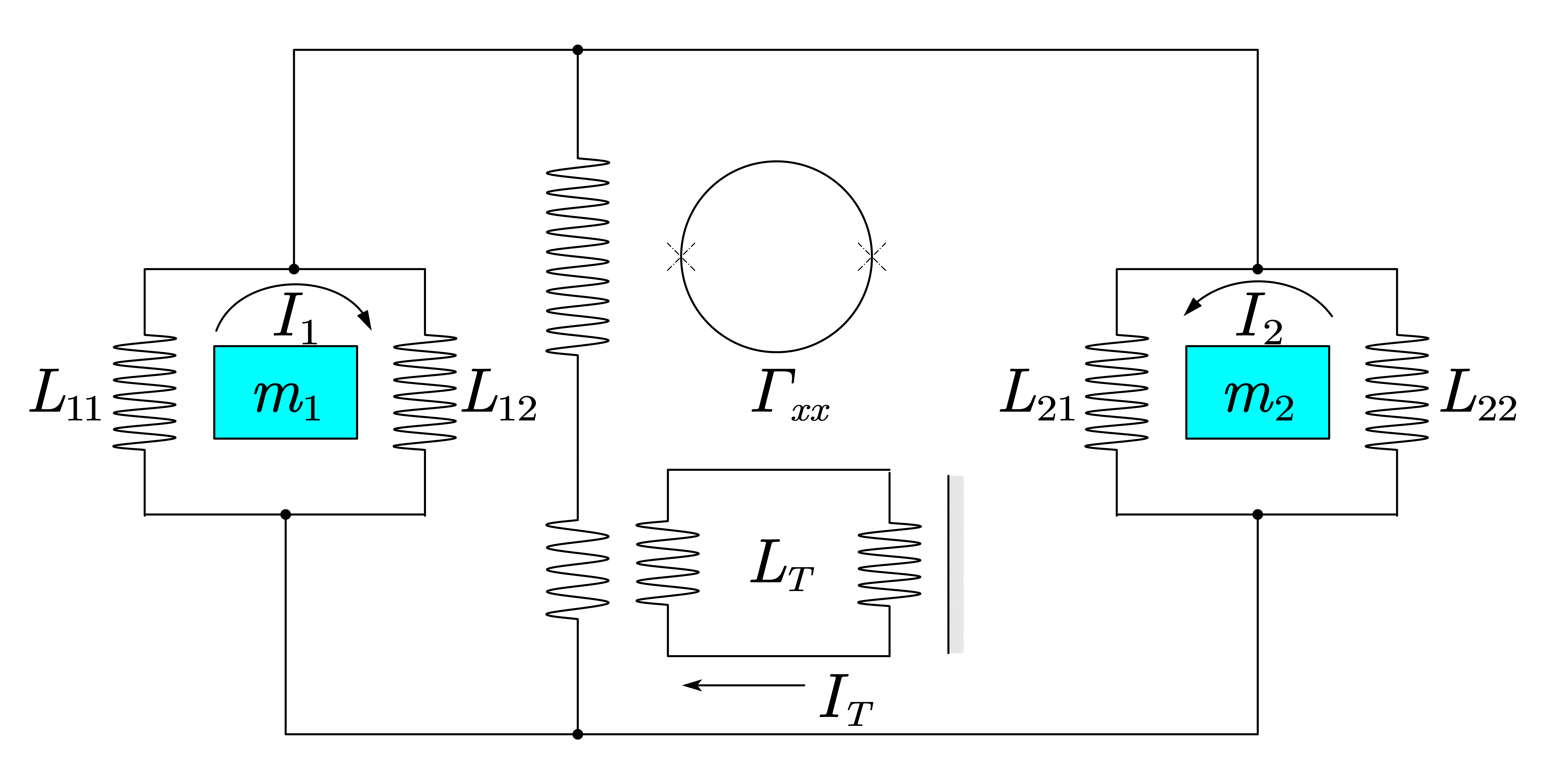}
    \caption{The principle of superconducting differential acceleration sensing~\cite{chan1987superconducting}}
    \label{superconducting}
\end{figure}

In the baseline design of highly sensitive gravity gradiometers operating in microgravity or zero-g environments in space, electrostatic or superconducting devices are commonly utilized. In these systems, pairs of proof masses are typically aligned along each measurement axis at a distance of approximately 50 cm. To mitigate disturbances affecting the proof masses, a range of strategies for isolation, position sensing, and control combinations are implemented, as extensively reviewed in~\cite{Moody2002, schumaker2003disturbance, Rummel2011}. A superconducting gravity gradiometer (SGG) consists of two (or more) superconducting accelerometers, each comprising a superconducting TM, a superconducting sensing coil carrying persistent sensing current, and a SQUID. Gravity (or other forces) moves the TM, modulating the inductance of the mass-coil system due to the Meissner effect. According to flux quantization, the sensing current adjusts to maintain conservative flux, generating a current signal detected by the SQUID~\cite{chan1987superconducting}. To measure gravity gradients and reduce common-mode noise such as satellite drag and temperature fluctuations, the two TMs are coupled with a superconducting sensing circuit shown in Figure.\ref{superconducting}.

\begin{figure}[!htbp]
    \centering    
    \includegraphics[height=4.7cm]{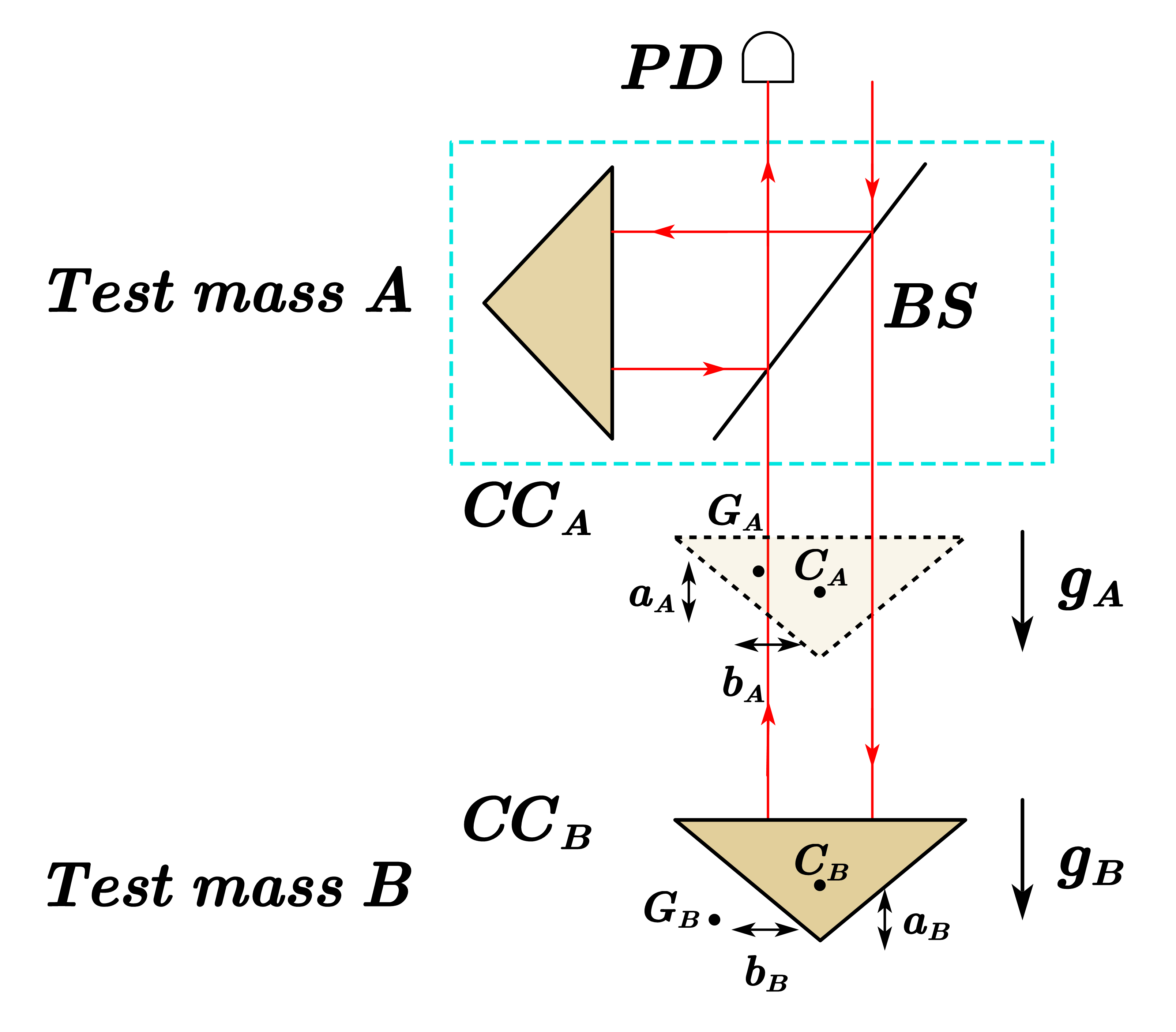}
    \caption{The principle of laser interferometer~\cite{shiomi2024rotational,shiomi2012development}}
    \label{laser}
\end{figure}

Figure \ref{laser} depicts the schematic optical layout of the laser interferometers~\cite{shiomi2012development}. Two TMs, \( A \) and \( B \), are simultaneously dropped from different heights within a vacuum chamber and experience gravitational fields \( g_A \) and \( g_B \), respectively. A laser beam aligned vertically is directed onto a cube beam splitter (BS) embedded in TM \( A \). The BS splits the laser beam into two arms: one reflects off a corner-cube prism \(CC_A\) embedded in test mass \( A \), and the other reflects off a corner-cube prism \(CC_B\) embedded in TM \( B \). These reflected beams are recombined at the BS, forming interference fringes based on relative optical-path changes. The interference fringes are detected by a photodetector (PD) located outside the vacuum chamber. The difference in free-fall acceleration (\( \Delta g = g_B - g_A \)) between the two TMs is determined by analyzing these interference fringes~\cite{shiomi2012development}. Solid glass corner-cube prisms are preferred for Laser Interferometric Gravity Gradiometers (LIGGs) due to their availability and ease of embedding within the test masses~\cite{shiomi2012development}.

In this article, we consider a single-satellite mission in Earth orbit. The satellite platform is an ultra-stable and ultra-quiet platform with drag-free control. Its payload includes a dual-axis high-precision laser interferometer for measuring gravitational gradients and a high-precision star-to-ground time-frequency comparison system. The satellite orbits in a 1500 km altitude Sun-synchronous circular orbit, maintaining a nadir-pointing attitude towards Earth. Utilizing data from the gradient meter, the satellite achieves two-degree-of-freedom drag-free control along the flight and orbit normal directions, ensuring the gravitational gradient instrument operates optimally.

Here we will constrain our parameter in details. For an orbital altitude of 1500 km, the orbital angular frequency is \( \omega = 9 \times 10^{-4} \) rad/s. We consider the gradiometers with a sensitivity of about 0.1 mE, which corresponds to an orbital frequency \( f = \sqrt{GM/a^3} = 0.144 \) mHz. Over one year's accumulation, the total cycles in the signal data amount to \( 4.5 \times 10^3 \). Therefore, for gradiometers with sensitivity better than \( 10^{-2} \) mE/\(\sqrt{\text{Hz}}\) in the low-frequency band near \( 0.1 \) mHz~\cite{griggs2017sensitive}. We could implement a proper and narrow bandpass filter, optimized for the signal frequency, to effectively eliminate unwanted noise and errors. Given that the signal is periodic, a 1-year dataset could then potentially reveal the noise floor at around \( 10^{-5} \) mE. If we set a suitable signal-to-noise ratio threshold for the 1 year measurement of the secular signal, by the displacement between test masses due to noise during the test, we expect that the constraint on the parameter \( \eta \) can be reached \(  \left | \eta - 1 \right |=\left | \frac{G}{G_N}-1 \right |=0.0025  \). For the optical gradiometers based on the new generation of space gravity gradiometer missions, constraints on such parameters may result in similar or even better bound.

\section{Conclusions}
\label{sec5}

In this paper, we have explored the weak-field and slow-motion limit of a covariant version of Horava-Lifshitz theory to constrain its parameters against recent space experiment results. This approximation is particularly effective in describing gravitational fields around Earth. It is anticipated that future gravity gradient measurement missions will provide more accurate data, enabling more precise constraints and scrutiny of modified gravity theories like Horava-Lifshitz Gravity that extend beyond Einstein's general relativity.

Our analysis demonstrates that high-precision laser gradiometers have the potential to improve the precision of parameter constraints by one or two orders of magnitude. That is to say, for a gravitational gradient detection mission with an orbital altitude of 1500km and an expected mission period of 1 year, the use of a new generation of equipment with better accuracy than  \( 10^{-2} \) mE/\(\sqrt{\text{Hz}}\) at the predetermined detection frequency can provide stronger constraints on modifying the gravitational theory. Take the Horava-Lifshitz theory in this paper as an example, the constraint on parameter \(\left|1-\frac{G}{G_N}\right|\) can be reached \(0.0025\). This is a significant advancement over previous methods, such as the Gravity Probe B (GP-B) experiment and the LAGEOS satellites, which have already provided valuable constraints on the parameters of Horava-Lifshitz theory. The innovative method we propose, utilizing laser interferometry in space, offers a new avenue for precise constraint of parameters such as G, the gravitational constant in Horava-Lifshitz theory.

The proposed method of measuring the GM field through differential measurement of the relative motion of two test masses in space is a promising approach. It allows for the direct measurement of the tidal force generated by the GM field, which is a part of the spacetime curvature. This method is distinct from previous efforts that attempted to track the orbital plane precession of a satellite through variations in the Keplerian elements with respect to the geocentric frame.

Furthermore, our work establishes a new theoretical framework for the differential measurement of the planetary GM field using a space-borne gradiometer. This framework is crucial for the design and implementation of future space missions aimed at measuring gravitational gradients with unprecedented precision. The potential of such missions is immense to test and constrain theories of gravity, including Horava-Lifshitz gravity.

In conclusion, the combination of advanced measurement techniques and theoretical frameworks presented in this paper opens new possibilities for the study of gravitational phenomena on a planetary scale. The pursuit of these avenues will not only refine our understanding of gravity but also contribute to the broader quest for a unified theory of quantum gravity.

\end{document}